\documentclass[a4paper,oneside,final,notitlepage,onecolumn,12pt]{article}

\usepackage[utf8]{inputenc}
\usepackage[english]{babel}
\usepackage{ifpdf,epsfig,array,amsmath,amssymb,psfrag}
\usepackage{ulem} 
\ifpdf
\usepackage[colorlinks=true,
filecolor=blue,linkcolor=blue,citecolor=blue,urlcolor=blue
]{hyperref}
\usepackage[usenames,dvipsnames]{color}
\else
\usepackage[usenames,dvips]{color}
\fi

\DeclareFontFamily{OT1}{rsfs}{} \DeclareFontShape{OT1}{rsfs}{m}{n}{
<-7> rsfs5 <7-10> rsfs7 <10-> rsfs10}{}
\DeclareMathAlphabet{\mycal}{OT1}{rsfs}{m}{n}

\DeclareFontFamily{OT1}{rsfs}{} \DeclareFontShape{OT1}{rsfs}{m}{n}{
<-7> rsfs5 <7-10> rsfs7 <10-> rsfs10}{}
\DeclareMathAlphabet{\mathscr}{OT1}{rsfs}{m}{n}

\begin{document}

\title{Gravitational radiation and isotropic change of the spatial geometry}

\author{\small 
Istv\'{a}n R\'{a}cz\thanks{email: iracz@rmki.kfki.hu}
\\ 
\small RMKI\\ 
\small  H-1121 Budapest, Konkoly Thege Mikl\'os \'ut 29-33.\\ 
\small Hungary
}
\maketitle

\begin{abstract} {To simplify a number of considerations in the weak field
    approximation, including the  determination of the response of
    interferometric gravitational wave detectors, the ``transverse traceless''
    (TT) gauge is often used. While the identification of the corresponding
    gauge invariant  part of the metric perturbations in the pure vacuum case
    is obvious, it is not widely known that the determination and the use of
    the TT part is much more complicated and, in turn, much less rewarding
    when sources are involved. It is shown here that likewise the transverse
    part of the electric current in the Coulomb gauge within Maxwell's theory
    the sources of the TT gauge part of the metric perturbations become
    non-local. This, in practice, invokes the necessity of the use of more
    adequate projection operators then the ones applied, e.g, in the weak
    field limit, and in many post-Newtonian considerations. It is also pointed
    out that, whenever nonlinear effects are taken into account, some of the
    conclusions concerning the response of interferometric gravitational wave
    detectors may be influenced. In particular, attention is called on the
    possibility that gravitational radiation may produce an isotropic change
    of the spatial geometry.} 
\end{abstract}


\vfill\eject

\section{Introduction}\label{int}

General relativity is a metric theory of gravity which can also be formulated
as being a gauge theory. This paper---besides pointing out various analogies
between the use of the Coulomb gauge in Maxwell's theory and the TT gauge in
the linearised Einstein's theory---is to point out the necessity of careful
reinvestigation of the standard arguments applied in determining the response
of interferometric gravitational wave (GW) detectors. The main motivation for
the present work is rooted in the fact that sensitivity of the ground based GW
detectors such as LIGO and Virgo has been improved significantly
\cite{ligo1,ligo2,virgo1,virgo2}, and it is widely held that, if not earlier,
then once the advanced detectors will be operating, detection of gravitational
waves will become an everyday routine. Therefore, it is getting of obvious
interest to have the best possible estimates for the astrophysical parameters
of the associated GW sources. In this respect it is worth to be recalled that
the optical observations of the change of the parameters of the orbital motion
of potential GW sources agree---see, e.g., the reports on the  Hulse-Taylor
pulsar and similar type of binary systems \cite{taylor,damour,csiro} (see also
section 6.2.3 of \cite{maggiore})---, up to a very high precision, with the
predictions of Einstein's theory. More precisely, the energy loss, which is
signified by the observed change of the orbital parameters and which is
assumed to be yielded by gravitational radiation, is in accordance with the
predictions of Einstein's theory. To ensure the same order of precision in
determining the astrophysical parameters based on the independent GW
observations it is of crucial importance to be sure that not only the
generation but the propagation of GW signals from the sources to our detectors
is properly determined in the applied models.  

\bigskip

In this paper we intend to provide a simple enough discussion 
indicating some of the potential sources of imprecision in the current
determination of the detector response for the arrival of GWs. In singling out
an appropriate framework it turned out to be really useful to have the
comprehensive paper by Flanagan and Hughes \cite{flanagan} at hand. Most of
the arguments below are going to refer to the results formulated in the first
part of this work. There is, however, one point to be mentioned here. As
opposed to the ambitious plans manifested by the first part of \cite{flanagan}
for certain reasons Flanagan and Hughes did not follow the path chosen
there. For instance, after providing a very useful critical summary of the
former conventional discussions in section 2.3 they returned to the orthodox
arguments in spite of the fact that these are apparently inconsistent with the
conclusion of the first part of their paper. In this respect, the main points
of the arguments presented in this paper may be considered as natural
continuation and completion of the work initiated by Flanagan and Hughes.

\bigskip

As it was indicated above the use of the Coulomb gauge in Maxwell's theory
shares several essential properties with that of the TT gauge in the weak
field approximation---in particular, since the argument justifying the
non-locality of the pertinent sources in these gauges are completely
parallel---in the rest of this section some of the most important related
facts of Maxwell's theory are recalled. The gauge invariant Maxwell tensor
$F_{\alpha\beta}$ is given in terms of a vector potential $A_\alpha$ as
$F_{\alpha\beta}= \partial_\alpha A_{\beta}-\partial_{\beta} A_{\alpha}$, and
two vector potentials $A_\alpha$ and $A'_\alpha$ are known to be physically
equivalent, i.e., they yield the same Maxwell tensor, if there exists a real
function $\chi$ such that 
\begin{equation}\label{vp}
A'_\alpha=A_\alpha + \partial_\alpha \chi\,.
\end{equation}
The field equations, whenever the gauge dependent vector potential $A_\alpha$
satisfies the Lorentz gauge condition $\partial^\alpha A_\alpha=0$,  read as
(see, e.g., \cite{jackson,wald}) 
\begin{equation}\label{feem}
\Box\,{A_{\alpha}} =-4\pi\, J_{\alpha}\,
\end{equation}
where $\Box=-\partial_t^2+\nabla^2$ and $J_{\alpha}$ stands for the electric
four current vector. It is known that by choosing the real function $\chi$
appropriately the Lorentz gauge condition can always be guaranteed to
hold. Recall also that we still have the freedom of  applying a restricted
gauge transformation of the form (\ref{vp}) provided that the generator $\chi$
is subject to the equation $\Box\,\chi=0$ since then the Lorentz gauge
condition remains intact.

Is is also well-known that gauge independent quantities can be built up from
the vector potential. The pertinent gauge is frequently referred as
``Coulomb'', ``radiation'' or ``transverse'' gauge and it can be introduced as
follows. Start by picking out an inertial reference system, $(t,{\bf x})$, of
the underlying Minkowski spacetime. Then a vector potential $A_\alpha$ may be
decomposed\,\footnote{%
The minus sign in front of the scalar potential
  comes from the fact the for a vector potential $A_\alpha$, with $A_0=-\phi$,
  entered into the formalism much later then the Coulomb potential.} as
$A_\alpha=(-\phi,A_i)$. The spatial part  $A_i$ of $A_\alpha$ can be split up
into `transversal' and `longitudinal' part as $A_i=A_i^T+\partial_i\varphi$,
where $A_i^T$ is such that $\partial^iA_i^T=0$. This decomposition is unique
if in addition the potential $\varphi$ is guaranteed to tend to zero while $r
\to \infty$ as then the elliptic equation $\nabla^2 \varphi=\partial^iA_i$
possesses a unique solution.  It is straightforward to see that once a gauge
transformation (\ref{vp}) is applied the variables $\phi$ and
$\varphi$ will be changed. It is well-known, however, that their combination
$\Phi=\phi+\partial_t\varphi$, along with $A_i^T$, is gauge invariant
\cite{jackson}. The field equation (\ref{feem}), pertinent for $\Phi$ and
$A_i^T$, reads then as 
\begin{eqnarray}
&&\nabla^2\,\Phi=-4\pi\, \rho\,\label{feem2}\\
&&\Box\,{A_{i}^T} =-4\pi\, J_{i}^T\,\label{feem3}
\end{eqnarray}
where the decomposition $J_\alpha=(-\rho,J_i)$ has been used. Notice that the
source for the transverse part of the vector potential $A_i^T$, 
\begin{equation}\label{tc}
J_{i}^T=J_{i}-\frac{1}{4\pi}\partial_i(\partial_t\Phi)
\end{equation}
extends over all space even if the spatial part $J_{i}$ of $J_{\alpha}$ is
localised \cite{jackson}. This is a direct consequence of the fact that $\Phi$
is subject to 
the Poisson equation (\ref{feem2}), i.e., $\Phi$ is non-local. Notice that
$\Phi$ is  time dependent even though it does not time evolve as a wave.

An immediate consequence of the non-locality of the transverse part of spatial
vector fields is that, once sources are involved, one has to be careful in
determining, e.g., the transverse part $A_i^T$ of the spatial part of vector
potential $A_i$ by making use of a projection operator. A projection
operator\,\footnote{  All the index raising and lowering are meant to be done
  by either of the fixed background metrics $\eta_{\alpha\beta}$ or
  $\delta_{ij}$ of the Minkowski spacetime or the Euclidean space,
  respectively. Moreover, the Einstein's summation convention is used only for
  identical upper and lower indices.}  ${P_i}^j$ of this type is formally
defined referring to the inverse Laplace operator, $\frac{1}{\nabla^2}$, as
${P_i}^j={\delta_i}^j-\partial_i\frac{1}{\nabla^2}\partial^j$. However, the
precise form of this projection operator---see, e.g., the discussion on page
242 of Jackson's book \cite{jackson}---, making the non-locality of the
involved fields completely transparent, assigns to a spatial vector $V_i$ its
transverse part $V_i^T={P_i}^jV_j$, as 
\begin{equation}\label{proj}
{P_i}^jV_j={\delta_i}^jV_j+ \frac{1}{4\pi}\partial_i^{^{\bf
    x}}\int\frac{\partial^j_{^{{\bf 
      x}'}} V_j({\bf x}') }{ |{\bf x}-{\bf
    x}'|}  d^3{ x}' \,,
\end{equation}
where the relation $\nabla^2_{\bf x}\,\frac{1}{|{\bf x}-{\bf
    x}'|}=-4\pi\,\delta({\bf x}-{\bf x}')$ has implicitly been used.

\medskip

It is straightforward to verify that whenever the spatial vector $V_i$
possesses the form of a plane wave solution, i.e.,  $V_i=V_i^0\cdot\cos({\bf
  k}\,{\bf  x}-\omega\, t+\psi_0)$,  with constant amplitude $V_i^0$ and phase
$\psi_0$, and with $\omega=|{\bf k}|$, then  ${P_i}^j$ can be given as 
\begin{equation}\label{planw}
{P_i}^j ={\delta_i}^j-n_i n^j\,,
\end{equation}
where the spatial unit vector $n_i$ is given as $n_i=k_i/{\omega}$. It is
important to emphasise that the projection operator ${P_i}^j$, given by
(\ref{proj}) for the generic case, reduces to the form of (\ref{planw}) {\it
  if and only if} $V_i$ is given as a linear superposition of plane wave
solutions such that all the spatial wavenumber vectors are
parallel. Accordingly, the application of the projection operator ${P_i}^j$
(\ref{planw}) does not yield the TT-part of $V_i$ besides this exceptional
case. It is worth to be mentioned that it does not even do the job for
slightly more general solutions to the sourceless wave equation,
$\Box\,V_i=0$.  All these observations imply then that whenever sources are
involved  the only adequate projection operator must possess the form of
(\ref{proj}).  

\section{The weak field approximation of GR}

The weak field approximation of general relativity is believed to be adequate
in describing weak gravitational effects. In such a case the metric
$g_{\alpha\beta}$ of the spacetime is supposed to be close  to the flat metric
$\eta_{\alpha\beta}$ of the Minkowski spacetime. More precisely, it is assumed
that Minkowski-type coordinate systems exist such that  
\begin{equation}\label{gytk}
g_{\alpha\beta}=\eta_{\alpha\beta}+h_{\alpha\beta}
\end{equation}
and that $|h_{\alpha\beta}|\ll 1\,$.

As a direct consequence of the generic diffeomorphism invariance of Einstein's
theory two linear perturbations $h_{\alpha\beta}$ and $h'_{\alpha\beta}$ of
the flat Minkowski spacetime are considered equivalent, whenever they are
related as
\begin{equation}\label{ge}
h'_{\alpha\beta}=h_{\alpha\beta}+\partial_\alpha\xi_\beta+\partial_\beta\xi_\alpha\,,
\end{equation}
where $\xi^\alpha$ denotes some infinitesimal vector field determining
the coordinate transformation 
\begin{equation}\label{ktrrrr}
x^a \rightarrow x'^a=x^a-\xi^a\,.
\end{equation}
Notice that in (\ref{ge}) $h_{\alpha\beta}$
and $\xi^\alpha$ play the same role as the vector potential $A_\alpha$ and the
function $\chi$ do in the Maxwell case.

The linearised Einstein equations can then be shown to take---in terms of the
trace reversed, 
\begin{equation}\label{tm}
\bar h_{\alpha\beta}=h_{\alpha\beta}-\frac12\,
\eta_{\alpha\beta}\,{h^\gamma}_\gamma\,,
\end{equation} 
of $h_{\alpha\beta}$---the simple form 
\begin{equation}\label{box} 
\Box{\,\bar h_{\alpha\beta}} =-16\pi\, T_{\alpha\beta}\,
\end{equation} 
provided that $\bar h_{\alpha\beta}$ satisfies the Lorentz gauge
condition  
\begin{equation}\label{LG}
\partial^{\alpha}{\bar h}_{\alpha\beta}=0\,.
\end{equation} 
It is well-known but worth to be mentioned that  there always exist 
coordinate transformations of the form $x'^\alpha=x^\alpha-\xi^\alpha$ such
that (\ref{LG}) holds in the new gauge.  Moreover, the pertinent gauge is not
unique since further restricted coordinate transformations with $\xi^\alpha$
subject to the wave equation  
\begin{equation}\label{ge3}
\Box \xi^\alpha=0\,
\end{equation}
may still be applied as they leave the Lorentz gauge condition (\ref{LG})
intact.  

The solution to the inhomogeneous equation (\ref{box}), given in terms of the
retarded Green function, read as 
\begin{equation}
\bar h_{\alpha\beta} (t,{\bf x})= 4 \int
 \frac{T_{\alpha\beta}
   (t-|\bf{x}-{\bf x}'|,{\bf x}')}   
 {|\bf{x}-\bf{x}'|} d^3{\bf x}'  \,.
\end{equation}
In virtue of (\ref{box}) all the components of $\bar
h_{\alpha\beta}$ possess radiative degrees of freedom which by many authors
(see, e.g., \cite{flanagan,hughes}) is considered to be an ``unfortunate
consequence'' of the applied gauge. A more adequate objection could be that
the components of $\bar h_{\alpha\beta}$ are not gauge invariant thus they
cannot be directly applied in determining the response of our GW detectors. 

\section{The  ``radiation'' or TT gauge}

In virtue of the criticism recalled above, more importantly, because of the
obvious need for a correct derivation of the response of our ground based
laser interferometric detectors like LIGO and Virgo to the arrival of a GW
signal, it is important to know whether the true radiative physical degrees of
freedom can always be separated in the weak field approximation. 

It has been known for long that the gauge independent expressions can be built
up from the components of $h_{\alpha\beta}$. In the following short review of
the pertinent results we shall follow the discussion of \cite{flanagan} unless
otherwise indicated.   

Consider first a ``$1+3$'' decomposition 
\begin{equation}
h_{\alpha\beta}=\left(
  \begin{tabular}{c|c}
        \text{ $h_{tt}$ }    &    \text{ $h_{ti}$ }     \\
        \hline
         \text{ $h_{it}$ }   & \text{ $h_{ij}$ }   \\
  \end{tabular}
\right)\label{dch}
\end{equation}
of $h_{\alpha\beta}$ based on the use of a Minkowski type coordinate system,
$(t,{\bf x})$, where {\it time-time}, {\it time-space} and {\it space-space}
parts are given in terms of the variables $\phi$, $\beta_i$, $\gamma$,
$\varepsilon_i$, $h_{ij}^{\rm TT}$ and $\lambda$ as 
\begin{eqnarray}
&& h_{tt} = 2\phi      \\
&& h_{ti} = \beta_i + \partial_i\gamma   \\
&& h_{ij}= h_{ij}^{\rm TT} + \frac{1}{3}H\,\delta_{ij} +
\partial_{(i}\varepsilon_{j)} + \left(\partial_i\partial_j -
\frac{1}{3}\delta_{ij}\nabla^2\right)\lambda\,,\label{H}
\end{eqnarray}
where $H \equiv \delta^{ij} h_{ij}$ denotes the three-dimensional trace which
is related to $h={h^\alpha}_{\alpha}$ as $h =H-2\phi$.  The variables $\gamma,
\varepsilon_i, \lambda$---and, in turn, $\beta_i$ and $h_{ij}^{\rm TT}$, as
well---can be                                               seen\footnote{%
  For more details the reader may look up the pertinent part of the argument
  applied for the  analogous decomposition of the energy-momentum tensor
  below.  } to be uniquely determined once the relations 
\begin{equation}\label{ellh}
\partial^i\beta_i = 0\ , \ \
\partial^i\varepsilon_i = 0\ , \ \
\partial^i h_{ij}^{\rm TT} = 0\,, 
\end{equation}
along with the boundary, or fall off, conditions 
\begin{equation}
\gamma \to 0,\ \ \varepsilon_i \to 0,\ \ \lambda \to 0,\ \
\nabla^2 \lambda \to 0\ \ {\rm while} \ \ r \to \infty\,,
\end{equation}
are imposed. Note that, in virtue of (\ref{H}) and (\ref{ellh}) the
contraction  $\delta^{ij}h_{ij}^{\rm TT}$ vanishes, which along with the last
relation of (\ref{ellh}), implies that $h_{ij}^{\rm TT}$ is TT.

As the components of $h_{\alpha\beta}$ themselves are not gauge invariant the
variables $\phi, \gamma, \lambda,$ $H, \beta_i$ and $\varepsilon_i$ are not
gauge invariant either. However, the combinations 
\begin{eqnarray}
&& \Phi \equiv -\phi + \partial_t\gamma - \frac{1}{2}\partial_t^2\lambda\\
&& \Theta \equiv \frac{1}{3}\left(H - \nabla^2\lambda\right)\\
&& \Xi_i \equiv \beta_i - \frac{1}{2}\partial_t\varepsilon_i\,,
\end{eqnarray}
along with the $3 \times 3$ matrix $h_{ij}^{\rm TT}$, can be shown to be gauge  
invariant. 

\subsection{The decomposition of the energy-momentum tensor}

In order to be able to determine the evolution equations for the above
introduced gauge invariant expressions we shall need an analogous
decomposition of the energy-momentum tensor. 

Before providing this decomposition recall first that whenever matter fields
are involved the Einstein's and matter field equations have to be solved
simultaneously. Now, partly to simplify our argument, and also to avoid the
associated considerable technical difficulties, without choosing any concrete
field equations we shall assume that the field values, along with the
components of the energy-momentum tensor, $T_{\alpha\beta}$, are determined by
some unspecified field equations, governing the time evolution of the sources
of the gravitational waves.  Once we have the energy-momentum tensor,
$T_{\alpha\beta}$, a decomposition, completely analogous to the one applied
above for $h_{\alpha\beta}$, can be provided as follows. 

Start by a ``$1+3$'' splitting of $T_{\alpha\beta}$
\begin{equation}
T_{\alpha\beta}=\left(
  \begin{tabular}{c|c}
        \text{ $T_{tt}$}    &    \text{ $T_{ti}$} \\
        \hline
         \text{ $T_{it}$}   & \text{ $T_{ij}$} \\
  \end{tabular}
\right) \,,
\end{equation}
and by defining the variables $\rho$, $S_i$, $S$, $\sigma_{ij}$, $\sigma_i$ and
$\sigma$ via the relations  
\begin{eqnarray}
&& T_{tt} = \rho \label{Ttt}\\
&& T_{ti} = S_i + \partial_i S \label{Tst}\\
&& T_{ij} = \sigma_{ij} + P\delta_{ij} + \partial_{(i}\sigma_{j)} +
\left(\partial_i\partial_j -
\frac{1}{3}\delta_{ij}\nabla^2\right)\sigma\,,\label{Tss}
\end{eqnarray}
where $P=\frac13\delta^{ij}T_{ij}$\,. As above $S$, $\sigma$ and $\sigma_i$
get to be uniquely determined once the relations 
\begin{equation}\label{dct}
\partial^i S_i = 0\ ,\ \
\partial^i\sigma_i = 0\ ,\ \
\partial^i\sigma_{ij} = 0\,,
\end{equation}
along with the boundary, or fall off, conditions
\begin{equation}
S \to 0,\ \  \sigma_i \to 0,\ \ \sigma \to 0,\ \
\nabla^2 \sigma \to 0\ \ {\rm while} \ \ r \to \infty \,,
\label{bct}
\end{equation}
are imposed. Note that as above, in virtue of (\ref{Tss}) and (\ref{dct}) the
contraction  $\delta^{ij}\sigma_{ij}$ vanishes, which along with the last
relation of (\ref{dct}), implies that $\sigma_{ij}$ is TT.

\medskip

The uniqueness of the above decomposition can be seen as follows.  First, the
$\partial^i$-divergence of  (\ref{Tst}) yields $\nabla^2 S = \partial^i
T_{ti}$, which has a unique solution by the above boundary condition. Once $S$
is known $S_i$ gets to be uniquely determined by the relation $S_i=T_{ti}
-\partial_i S$. Concerning the uniqueness of $\sigma$, take now the
$\partial^i \partial^j$-``divergence'' of (\ref{Tss}) which yields the Poisson
equation $ \nabla^2 \nabla^2  \sigma = \frac32 \left[\partial^i \partial^j
  T_{ij} - \nabla^2 P\right]$, and which has a unique solution for
$\nabla^2\sigma$, and, in turn, in virtue of (\ref{bct}), $\sigma$ becomes
uniquely determined, as well. Once $\nabla^2\sigma$ is known $\sigma_i$ gets
also to be uniquely determined by the Poisson equation  $\nabla^2 \sigma_i =
2\,[\partial^j T_{ij} -  \partial_i P] - \frac43\, \partial_i \nabla^2
\sigma$, which is yielded by the $\partial^j$-divergence of (\ref{Tss}), along
with the fall off condition $\sigma_i \to 0$ while $r \to \infty $. 

\medskip

As the energy momentum tensor is supposed to be known it is straightforward to
see that having $\sigma$ and $\sigma_i$ determined, 
$\sigma_{ij}$ gets also fixed as 
\begin{equation}\label{sigmij}
\sigma_{ij} = T_{ij}- P\delta_{ij} - \partial_{(i}\sigma_{j)} -
\left(\partial_i\partial_j -
\frac{1}{3}\delta_{ij}\nabla^2\right)\sigma\,. 
\end{equation} 
In \cite{flanagan} the authors claim that $\sigma_{ij}$, along with some of the
other variables, can be chosen freely. It should be noted that this can be done
only if the energy-momentum tensor is not specified. 

There is an even more important additional point to be mentioned here. 
As the variables
$\sigma$ and $\sigma_i$ satisfy Poisson type equations they are non-local. In
consequence of this fact and the above relation (\ref{sigmij}) the TT-part,
$\sigma_{ij}$, of $T_{ij}$ cannot be local either regardless whether the
energy-momentum tensor, $T_{\alpha\beta}$, of the matter sources is of compact
support or not.  

Since this non-locality is in certain extent inconvenient it could be tempting
to argue that although the variables  $S$, $\sigma$ and $\sigma_i$ were shown
to be subject to some very complicated elliptic equations they might be
completely negligible. In this respect it is useful to take into account the
conservation law $\partial^\alpha T_{\alpha\beta} = 0$ which reads as
\cite{flanagan} 
\begin{eqnarray}\label{div}
&& \nabla^2 S = \partial_t\rho\\
&& \nabla^2 \sigma = -\frac{3}{2}\,P + \frac{3}{2}\,\partial_t S\\
&& \nabla^2\sigma_i = 2\,\partial_t S_i\,.
\end{eqnarray}

These relations---which had a completely different role in the discussion of
\cite{flanagan}---provide the following alternative characterisation of the
variables $S$, $\sigma$ and $\sigma_i$. They make it immediately transparent
that the time derivative of the energy density and that of the impulse---both
of these quantities are supposed to provide significant contribution to the
gravitational wave production---are the sources for $S$ and $\sigma_i$,
respectively.  Thereby, the more intensive is the considered GW source the
more significant the quantities $S$, $\sigma$ and $\sigma_i$ become. 

\section{The linearised Einstein's equations}

Now we are prepared to present the explicite form of Einstein's equations
relevant for the above introduced gauge invariant quantities in the
investigated weak field approximation. These equations can be given as
\cite{flanagan}  
\begin{eqnarray}
&&\nabla^2\Theta = -8\pi\rho \label{1}\\
&&\nabla^2\Phi = 4\pi\left(\rho + 3P - 3\,\partial_t S\right)\label{2}\\
&&\nabla^2\Xi_i = -16\pi S_i\label{3}\\
&&\Box\, h_{ij}^{\rm TT} = -16\pi\,\sigma_{ij}\,.\label{TT}
\end{eqnarray}

The above equations justify the conventional assertion that only the TT part
of the metric perturbation satisfies wave equation while all the other gauge
invariant expressions, although they are time dependent, do not time evolve as
waves since they are subject to Poisson equations. Accordingly, it is
frequently said that only the ``non-radiative'' physical degrees of freedom
are tied to the matter sources.  What is even more surprising is that
conclusions of the following type are drawn based on the above set of
equations: {\it Since the sources are at enormous distance from the Earth, in
  virtue of (\ref{TT}), GW signals can basically be considered as being
  sourceless and possessing the same type of properties as if they were GW
  signals in the pure vacuum case.}  As opposed to this, we would like to
emphasise that according to the conclusion of the previous section the source
term $\sigma_{ij}$ in (\ref{TT}) is non-zero, and it is non-local either even
though $T_{\alpha\beta}$ is completely localised.

\bigskip

As an immediate consequence of this non-locality let us make a comment
regarding  the ``conventional'' way of determining the TT part $h_{ij}^{\rm
  TT}$ of a solution $h_{\alpha\beta}$ to the evolution equation
(\ref{box}). It is usually assumed in the weak field approximation (see, e.g,
Section 4.1 of \cite{flanagan} or section 3.1 of \cite{maggiore}) and, for
some surprise, also in the post-Newtonian framework (see, e.g., the sentence
involving Eq.\,(2.2) of \cite{favata}, section 5.3.4 of \cite{maggiore} or the
first paragraph on page 19 in \cite{blanchet}), that the TT part of $h_{ij}$
may be determined in terms of the projection tensor, 
\begin{equation}\label{prl}
{\Lambda_{ij}}^{kl}={P_i}^k{P_j}^l-\frac12 P_{ij}P^{kl}\,,
\end{equation}
as $h_{ij}^{\rm  TT}={\Lambda_{ij}}^{kl}h_{kl}$,  where the ``elementary
projection operator'' ${P_i}^j$ is supposed to possess the form ${P_i}^j
={\delta_i}^j-n_i n^j$. We would like to emphasise here that, as it follows
from the argument outlined at the end of section\,\ref{int}, this form of
${P_i}^j$ cannot adequately be applied even to the superposition of plane wave
solutions to the  sourceless wave equation unless all the spatial wave number
vectors are parallel. Since the evolution equation (\ref{box}) for
$h_{\alpha\beta}$ must have sources in astrophysical situations, as actually
we wish to observe these sources, the projection operator
${\Lambda_{ij}}^{kl}$, when it is expressed in term of the ``elementary
projection operator'' via (\ref{prl})---as opposed to the generic resolution
applied in various calculations (see, e.g.,
\cite{flanagan,maggiore,favata,blanchet})---has to be constructed by making
use of (\ref{proj})  instead of applying (\ref{planw}). The corresponding
projection operator---the complexity of which is expected to reflect all the
technical difficulties related to the non-locality of $\sigma_{ij}$---will
produce the adequate TT-part of $h_{ij}$. In virtue of these observations
there is an obvious need for the reinvestigation of the procedures yielding
the wave forms in the asymptotic region by applying the correct projection
operator ${\Lambda_{ij}}^{kl}$, e.g, in the post-Newtonian framework.

\section{Further implications of non-localities}

In proceeding, let us recall now, that in many of the arguments, aiming to
determine the response of laser interferometric  detectors to the arrival of a
GW signal, the calculations end up (see, e.g., Eq.\,(3.10) of \cite{flanagan},
or Eq.\,(1.93) of \cite{maggiore}) with
the variant of the geodesic deviation equation 
\begin{equation}
\frac{d^2L^i(t)}{dt^2}=-{R^i}_{tjt}L^j\,, 
\end{equation}
where $L^i(t)=L^i_0+\delta L^i(t)$, with $\delta L^i(t)\ll L^i_0$ and
$i,j=1,2$, is supposed to denote the coordinates of mirrors at the end of  the
arms in the proper detector frame.  

Then, in the linearised theory, assuming that no sources are present, the
relation $R_{itjt} = - \frac{1}{2} {\partial_t^2 h}_{ij}^{\rm  TT}$, along
with the assumption that both  $\delta L^i(t)$ and $\partial_t\left(\delta
L^i(t)\right)$ vanish at $t=0$, is applied to derive the familiar ``gauge
independent'' relation 
\begin{equation}\label{dr1}
\delta  L_i(t)=\frac12{ h}_{ij}^{\rm TT}L^j_0\,.
\end{equation}

However, as it has already been emphasised above, since we do want to make
astrophysical observations, the presence of the sources has to be taken into
account. In  this more realistic situation, as opposed to the pure vacuum
case, the gauge invariant ``tidal force components'' of the Riemann tensor
read as (see, e.g., \cite{flanagan}) 
\begin{equation}\label{tidalf}
R_{itjt} = - \frac{1}{2} {\partial_t^2 h}_{ij}^{\rm TT} +
\partial_i\partial_j\Phi + \partial_t\partial_{(i} \Xi_{j)} - \frac{1}{2}\,
        {\partial_t^2  \Theta}\,\delta_{ij}\,. 
\end{equation}
It is well-known that in the linearised theory, if one takes into account the
conservation of the stress energy tensor,
$\partial^\alpha T_{\alpha\beta}=0$, the last three terms can be
shown to fall off like $1/r^2$ or faster \cite{flanagan}.\footnote{In virtue
  of (\ref{1})-(\ref{3}) $\Phi, \Theta$ and $\Xi_i$ fall of like
  $1/r$. However, the coefficients of the $1/r$ parts of these quantities can
  be given in terms the conserved mass $M=\int\rho\,{\rm d}x^3$ and the
  conserved linear momentum $P_i=\int S_i\,{\rm d}x^3$ \cite{flanagan}.}
Thus, the only term with $1/r$  fall off is the first term on the right hand
side of (\ref{tidalf}). Thereby, within the linearised theory regardless
whether sources are present relation (\ref{dr1}) determine the response of our
detectors. 

Up to now only the linearised theory has been considered. However, it has been
known for long (see, e.g., the discussion in section 4.4. of \cite{wald}) that
it has serious limitations, since, even to have a consistent Newtonian limit
one must go beyond the linear approximation. This is justified by the fact
that in the linear theory, in virtue of the conservation law $\partial^\alpha
T_{\alpha\beta}= 0$,  e.g., the members of a binary system---instead of
orbiting around each other---have to follow timelike geodesics, i.e., straight
lines, of the Minkowski spacetime. As opposed to this a nearly Newtonian type
of orbiting is produced by the binary if the terms higher order in
$h_{\alpha\beta}$ are restored in the Einstein tensor $G_{\alpha\beta}$.
Therefore, in a physically consistent description of gravitational wave
generation processes the back-reaction has to be taken into account. This can
be done, while preserving the simplicity of the basic equations of the linear
approximation, by putting all the non-linear terms of the Einstein tensor to
the energy-momentum tensor side, or, more precisely, by replacing in
(\ref{box})  the energy-momentum tensor $T_{\alpha\beta}$ by the sum
$T_{\alpha\beta}+t_{\alpha\beta}$, where $t_{\alpha\beta}=
-\frac{1}{8\pi}\hskip.05cm{}^{^{(nl)}}\hskip-.07cm G_{\alpha\beta}$, and
$\hskip.05cm{}^{^{(nl)}}\hskip-.07cm G_{\alpha\beta}$ consists of all the
higher order terms in the Einstein-tensor. Note that the assumption,
$|h_{\alpha\beta}|\ll 1\,$, requiring the perturbations  to be sufficiently
small, could also be relaxed then. In particular, once $T_{\alpha\beta}$ is
replaced by the sum $T_{\alpha\beta}+t_{\alpha\beta}$ the yielded equations
become suitable to describe the evolution of intrinsically strong GW sources
which cannot be done properly in the linearised theory.  Note also that the
conservation law $\partial^\alpha T_{\alpha\beta}= 0$, which is responsible
for the above mentioned defects, gets to be replaced
by the more adequate relation $\partial^\alpha
(T_{\alpha\beta}+t_{\alpha\beta}) = 0$.  Once the replacement
$T_{\alpha\beta}\rightarrow T_{\alpha\beta}+t_{\alpha\beta}$, along with a
senseful redefinition of the quantities $\rho$, $S_i$, $S$, $P$,
$\sigma_{ij}$, $\sigma_i$, has been done then all the previously derived
equations, (\ref{div})--(\ref{TT}), can be seen to preserve their forms. 

\medskip

Now, by making use of the above introduced nonlinear setup, we intend to
provide a plausibility argument suggesting that the response of our detectors
is going to be affected by back-reaction.  Before presenting our argument we
would like to emphasise that, whenever nonlinearities are taken into account
but sufficiently far from the sources $|h_{\alpha\beta}|\ll 1\,$ holds, it
seems to be reasonable to assume that the tidal forces can still be given by
(\ref{tidalf}), with the distinction that now $\Theta, \Phi, \Xi_i$ and ${
  h}_{ij}^{\rm TT}$ refer to the redefined quantities. We
shall use this assumption below. Note also that in
consequence of the hidden nonlinearities no attempt is made to go beyond
providing a plausibility argument, i.e., no quantitative estimates are
derived. 

In proceeding let us revisit the fall off properties of the last three terms
on the right hand side of (\ref{tidalf}).  Note first that since the middle
two terms in (\ref{tidalf}) contain spatial derivatives, it is straightforward
to verify that they decay faster than the terms on the sides. In particular,
since both $\Phi$ and $\Xi_i$ fall off like $1/r$ the relations
$\partial_i\partial_j\Phi\sim {r}^{-3}$ and $\partial_t\partial_{(i}
\Xi_{j)}\sim {r}^{-2}$ can be seen to hold. Therefore the main issue is
whether, besides ${ h}_{ij}^{\rm  TT}$, the last term on the right hand side
of (\ref{tidalf}) may also have a $1/r$ fall off in the nonlinear case.  In
this respect the following simple example provides some important clues. 

Assume that in our spacetime we have nothing else but a localised GW source
which produces a single short lasting GW burst. Consider now an observer that
is asymptotically far from the source in the distance. For simplicity let us
represent the world-sheet of the GW burst, as it is travelling from the source
towards infinity, by a null shell. The observation occurs where the world-line
of the observer meets this null shell. In advance to the observation the $1/r$
part of $\Theta$ (as measured by the observer) is proportional to the ADM
mass. However, if back-reaction is taken into account, regardless how tiny is
the energy carried by the GW on the future side of the null shell, the
coefficient of the $1/r$ part of $\Theta$ will be smaller. This is so because
the mass felt by the observer will be smaller than the ADM mass as the GW,
even after its detection, goes on, carrying the energy released by the source,
towards null infinity. Accordingly, the coefficient of the $1/r$ part of
$\Theta$ will vary in time in spite of the fact that the total ADM mass is
conserved.

The above example suggests that nonlinearities manifest themselves in the
following simple way. Whenever $T_{\alpha\beta}$ is replaced with
$T_{\alpha\beta}+t_{\alpha\beta}$---in consequence of the fact that
$t_{\alpha\beta}$ extends beyond the observer and it is time dependent---the
coefficient of the $1/r$ part of $\Theta$, at the location of the observer,
vary with time. If the observer is not too  far from the sources then this
time variation might be oscillatory although it is expected to be monotonic,
as in the above example, if the observer is asymptotically far from the
sources. 

Thereby, in the nonlinear regime it seems to be reasonable to assume that
second time derivative of both $\Theta$ and ${ h}_{ij}^{\rm  TT}$ fall off
like $1/r$. Accordingly, in virtue of (\ref{tidalf}), for the variation of the
coordinates of the end mirrors
the relation 
\begin{equation}\label{vpl}
\delta L_i(t)\approx\frac12\left[{ h}_{ij}^{\rm  TT}+{\Theta}
  \delta_{ij}\right]L^j_0
\end{equation}
applies. What is even more remarkable is that the effect of the second term
on the right hand side of (\ref{vpl}) on the arms of the laser interferometric
detectors is nothing but a possibly tiny but isotropic change of the arm
lengths.

Of course, without carrying out further quantitative investigations there is
no way to argue that this effect is important. Thus, it is of obvious interest to
know whether this effect is large enough, and whether it could be detected by
the current arrangements of our ground based laser interferometric
detectors. Unfortunately, the answer to the latter question is no since the
LIGO-Virgo type detectors in their present form are sensitive only to the
relative variation of the arm lengths. We hope that by a suitable modification
of the applied detection schemas, e.g., by adopting some of the ideas proposed
in \cite{tarabrin}, it may be  possible to measure the variation  of the arm
lengths separately. If this can be done an isotropic change of the spatial
geometry, simultaneous to the arrival of a gravitational wave train, could
hopefully be detected.  

\section{Final Remarks}

In this paper some of the peculiarities of the TT gauge in the weak field
approximation were investigated. The results found indicate that there have
been several assumptions applied in determining the TT part of the metric 
perturbations which are in the air. In particular, the results found have 
the following non-trivial consequences. 

\medskip

First, it is pointed out that, whenever sources are involved, in determining
the TT part of the metric perturbations a new approach---significantly
different from the currently applied one---, which is taking into account the
non-locality of the sources pertinent for the gauge independent TT
variables is needed. It is also indicated that the associated improvements
will affect the asymptotic wave forms not only in the linearised theory but
in the post-Newtonian framework, as well.   

Second, the determination of the response of interferometric gravitational
wave detectors may be influenced considerably if back-reaction is taken into
account. A plausibility argument was provided justifying that the arrival of a
GW train yields an additional isotropic change of the arm lengths. We would
like to emphasise that there is an immediate consequence of this effect which
may affect the estimates concerning the detectability of gravitational wave
signals by our current detectors. This is related to the possibility that the
energy released by the astrophysical sources may not completely be transferred
into the pure radiative degrees of freedom but some part of it could be used
to produce the isotropic change of the spatial geometry in the distance. This
variation of the volume may decrease the current estimates of  GW amplitudes
given in terms of ${ h}_{ij}^{\rm  TT}$. It is also indicated that the
anticipated isotropic change may be observed by suitably modified versions of
the currently applied laser interferometric detectors. 

\medskip

It is worth to be mentioned that in most of the investigations of GW
productions, whenever there is an attempt to take into account the nonlinear
back-reactions, curiously enough, the nonlinearities themselves are left out
almost immediately from the discussions. For instance, in determining the
quadrupole tensor, which, in the nonlinear case, should read as 
\begin{equation}\label{quad}
q_{\alpha\beta} = \int_\Sigma  \left[ T^{00}+ t^{00} \right]x^\alpha x^\beta
\,{\rm d}^3x
\end{equation}
it is usually assumed (see, e.g., the bottom of p.\,87 in \cite{wald}, or
section 4.2 in \cite{flanagan}) that the contribution of $t^{00}$ can be
neglected as the relation $t^{00} \ll T^{00}$ holds at the location of the
source. However, in accordance with the last remark of Wald on the top of
p.\,88 of \cite{wald}, we would like to emphasise that a higher level of
clarity and rigour should be involved here. Note that the main results of the
present paper support these necessities simply because $t^{00}$ is known to be
global, and, far from the sources, where $t^{00}$ may be small but positive,
its contribution could also be significantly amplified by the factor $x^\alpha
x^\beta$ in (\ref{quad}). Therefore, it may happen that whenever $t^{00}$ is
sufficiently anisotropic its contribution to the quadrupole moment tensor will
become significant. This, in turn, may yield  a considerable deformation of
the emitted wave forms while travelling from the sources to the observers. If
this turns out to be the case we cannot avoid a careful revision of the
currently applied template banks.

\medskip

It is also of obvious interest to know what might be the relation between
our findings and ``Christodoulou's nonlinear memory effect''
\cite{christodoulou}. In this respect we would like to mention that, because
of the significant differences of the mathematical setup applied in
\cite{christodoulou} and in this paper, it is not obvious at all to derive a
meaningful relation especially because in \cite{christodoulou} no explicite
expression is given that could represent the displacement of the mirrors. In
particular, as a consequence of these differences while in this paper the
nonlinear effects were shown to be isotropic no such conclusion was derived
in \cite{christodoulou}.  Nevertheless, it is worth to be mentioned here that
according to the rough estimate provided by Christodoulou the nonlinear
effects may be of the same order as the linear ones.

\medskip

It might be tempting to consider further consequences of the indicated
isotropic change of the spatial geometry. Indeed, the pertinent implications
might be far-reaching if at certain parts of the universe the monotonous
increasing of the function $\Theta$ could be verified. Nevertheless, we would
like to emphasise again that in this respect the present paper is far from
being conclusive as neither reliable estimates concerning the fraction of the
energy converted into the change of $\Theta$ has been derived nor its
monotonicity has been studied. To do so further analytic and numerical
investigations are needed.


\section*{Acknowledgments}  
The author wishes to thank Robert Wald and Andor Frenkel for useful
discussions, for reading the
manuscript and for suggesting a number of improvements. This research was
supported in parts by OTKA grant K67942.

\end{document}